\documentclass{elsart}
\usepackage{epsfig,wrapfig}
\begin{document}
\begin{frontmatter}
\title{\boldmath Image and Non-Image Parameters of 
Atmospheric Cherenkov Events: 
a comparative study of their $\gamma$-ray/hadron 
classification potential in UHE regime}
\author[MUMBAI]{A.~Razdan}\footnote{corresponding author; e-mail: 
akrazdan@apsara.barc.ernet.in}
\author[KARLSRUHE]{A.~Haungs}
\author[KARLSRUHE]{H.~Rebel}
\author[MUMBAI]{C.L.~Bhat}
\address[MUMBAI]{Bhabha Atomic Research Centre, Nuclear Research 
Laboratory, Mumbai-400 085, India}
\address[KARLSRUHE]{Forschungszentrum Karlsruhe, Institut f\"ur 
Kernphysik, Postfach 3640, D-76021 Karlsruhe, Germany}

\begin{abstract} 
In this exploratory simulation study, we 
compare the event-progenitor classification 
potential of a variety of measurable
parameters of atmospheric Cherenkov pulses which are
produced by ultra-high energy (UHE) $\gamma$-ray
and hadron progenitors and are
likely to be recorded
by the TACTIC array of atmospheric Cherenkov telescopes. 
The parameters derived from Cherenkov images include Hillas, 
fractal and wavelet moments, while
those obtained from non-image Cherenkov data consist of pulse 
profile rise-time and base width and the relative ultraviolet to 
visible light content of the Cherenkov event.  
It is shown by a neural-net approach that these parameters,
when used in suitable combinations, can bring about a
proper segregation of the two event types,
even with modest sized data samples of progenitor particles.
\end{abstract}


\end{frontmatter}

\section{Introduction}
Atmospheric Cherenkov telescopes, being extensively deployed for ground-based
$\gamma$-ray astronomy work in the TeV photon energy range for the last 30
years, have invariably to deal with source signals which are extremely
weak, both, in absolute terms as also in relation to the background cosmic-ray
events (typically $\le$ 1:100 at photon energies $\ge$ 1 TeV).  
This has made it mandatory to devise
and continually upgrade techniques, in both hardware and software, for
optimum acceptance of $\gamma$-ray events and maximal rejection of the
background events.  A path-breaking advance was made in this direction
through the successful implementation of the Cherenkov imaging technique,
first attempted by the Whipple collaboration \cite{whipple}.  
Here a Cherenkov event
is 'imaged' in the focal-plane of a Cherenkov telescope by recording
the two-dimensional distribution of the resulting light pattern with the
help of a Cherenkov imaging camera, generally consisting of a square matrix
of fast photomultiplier tubes (typical FoV $\sim 3^\circ -6^\circ$, pixel
resolution $\sim 0.15^\circ - 0.3^\circ$).  The recorded image, after necessary
pre-processing, is parameterised into 'Hillas' parameters -- mainly a
set of second-moments which include image shape parameters, called
Length (L) and Width (W) and image orientation parameters, like Azwidth (A)
and Alpha ($\alpha$) \cite{hillas}. 
Both simulation and experimental studies have shown that
$\gamma$-ray images are more regular and compact (smaller L and
W) as compared with their cosmic-ray counterparts and have a well-defined
major axis (orientation) which, in the case of $\gamma$-rays coming from
a point $\gamma$-ray source, are oriented closer towards the telescope
axis (smaller A and $\alpha$ vis-a-vis randomly oriented cosmic-ray images).
Following this approach, the presently operating Cherenkov 
imaging telescopes, operating in mono or
stereo-observation modes, have been able to reject cosmic-ray background
events at $\ge 99.5\%$ level, while retaining typically $\sim 50\%$ 
of the $\gamma$-ray
events from a point $\gamma$-ray source and thus register a substantial
increase in their sensitivity, compared with non-imaging,
generation-I systems.  \\
As a consequence, today, not only do we have
for the first time firm detections of at least half a dozen 
compact $\gamma$-ray
sources, but in several cases, their spectra in the TeV energy domain
are reasonably well delineated to permit a realistic 
modeling~\cite{aharonian1}.
Despite this spectacular success of the imaging technique, witnessed
during the last one decade, there is still a sense of urgency 
for seeking a further significant improvement in
the sensitivity level in the TeV $\gamma$-ray astronomy field.
There are 3 main reasons for this need: One,
the detection
of fainter compact sources, understandably expected to
constitute the major bulk of the TeV source 'catalogue'.
Secondly, the detection of $\gamma$-rays from non-compact 
sources or of a diffuse
origin.  Only the image shape parameters like L and W, can 
be used in this case.
They are known to be poorer event classifiers compared with the 
orientation parameter, $\alpha$, in the case of point $\gamma$-rays 
sources and are
found to be grossly inadequate to deal with, for example, the 
diffuse $\gamma$-ray
background of galactic or extragalactic origin.  The third motivation for
seeking better event-classifier schemes stems from the desirability 
to use Cherenkov
imaging telescopes in a supplementary mode of observations for
UHE $\gamma$-ray astronomy and cosmic-ray
mass-composition studies in tens of TeV energy range and thereby 
secure independent information on these important problems
through this indirect but effective
ground-based~\cite{fegan} technique.  Evidently, the required
classification schemes in
this case will need  to have the capability of not only segregating
$\gamma$-rays
from the general mass of cosmic-ray events, but also to act, at least, as a
coarse mass-spectrometer and separate various cosmic-ray
elemental groups \cite{haungs1}. Hillas parameters
are found to be very good classifiers for smaller images (close to telescope
threshold energy) but tend to fail for very large images (higher primary
energies of the $\sim$ 10's TeV) since too many tail pixels are included in
the image. \\
With the above-referred broad aims in mind, serious attempts are
presently on to seek enhanced sensitivities and event-characterization
capabilities for Cherenkov systems through the deployment of more efficient
or versatile image analysis techniques and the 
inclusion of non-imaging parameters in these classification schemes,
for example, rise-time and base-width of the recorded time
profile of the atmospheric Cherenkov event or its relative spectral
content, i.e. the ultraviolet (U) to visible (V) 
light flux ratio (U/V ratio).  
Following in the same spirit, we have recently shown in 
\cite{haungs1} that fractal parameters can be effectively used
to describe Cherenkov images. 
Thus, it seems  meaningful to supplement Hillas parameters
with appropriate fractal moments for seeking a better characterization
of these images w.r.t. progenitor particle type.  In fact, in \cite{haungs1},
it has been shown that, by exploiting correlations amongst 
Hillas parameters with fractal
and wavelet moment parameters of Cherenkov images with the help of a properly
trained artificial neural network, it is possible not only to efficiently
segregate $\gamma$-rays events from the general family of cosmic-ray events
but also to separate the latter into low (H-like), medium (Ne-like) and
high (Fe-like) mass-number groups with a fairly high quality factor -- a
job which Hillas parameters alone cannot do as efficiently under similar
conditions of net training. 
Recent work done on this subject by HEGRA \cite{hegra} 
group tends to support the above conclusions.
The main reasons why this multi-parameter
diagnostic approach works is that these parameters look at different aspects
of the Cherenkov image: while Hillas parameters are based on its geometrical
details, the fractal and wavelet moments are sensitive to image intensity
fine-structure and gradients.  In this first exploratory exercise, we worked
with a large data-base of 24,000 events (consisting of equal
numbers of $\gamma$-rays, protons, neon and iron nuclei) and successfully
sought their separation by using Hillas, fractal and wavelet-moment
parameters.  In the present feasibility study, we essentially
invert the classification
strategy by using an extremely small database (100 events each belonging to
$\gamma$-ray, proton, neon and iron parents) and seek their segregation
into two main parent species (photons and nuclei) by using a significantly
larger parameter-space, consisting of both image (Hillas, fractal and
wavelet) and non-image (time-profile and spectral)
classifiers.\\
The present study has a particular relevance for UHE $\gamma$-ray
emissions (10's of TeV) from point sources which are expected
to be extremely weak and will need more sensitive signal-retrieval strategies
than the one provided by the Hillas image-parameterization scheme alone.
The imaging element of the TACTIC array is especially geared for these
investigations for it has an unusually large field of view (FoV) of
$\sim 6^\circ\times 6^\circ$ with a uniform pixel
resolution of $\sim$ 0.31. It can
follow UHE events to larger impact parameter values (400-600~m) and
large zenith angle (typically $> 40^\circ$) and thus
expect to detect these events at a significantly higher
rate than would be possible
for imaging systems with a smaller FoV ($< 3^\circ$).
The related problem of
pixel saturation will also not arise generally because most of these UHE
events will belong to larger impact
parameters (typically $>$ 400~m), where the
Cherenkov photon density will be appreciably lesser
than what it is expected at smaller zenith angles and smaller 
extensive air shower (EAS) core
distances ($\le$ 150~m). It is also conceivable that, for at least some
$\gamma$-ray sources, the photon spectrum is significantly flatter
than what is known for the standard TeV $\gamma$-ray candle source,
Crab Nebula (differential photon number exponent $\sim$~2.7). In
such an eventually, one would expect to record a significantly
higher flux of UHE photons ($\ge$ 50~TeV) than what follows from 
a linear extrapolation of the known Crab spectrum in the TeV region.
The profound astrophysical implications following an unequivocal
detection of an UHE $\gamma$-ray sources $\ge$ 50~TeV) and the
unique promise  that the TACTIC imaging telescope offers for
such an investigation owing to its large FoV and uniform pixel
resolution are two factors which have  motivated us to carry
out the present evaluation exercise. \\
In this paper, we start with a description of the TACTIC array \cite{tactic},
particularly highlighting the salient features of relevance to the present
work.  This is followed by a discussion on the methodology adopted for
data-base generation for this instrument, using the CORSIKA simulation
code \cite{capdevielle,heck}
and the subsequent derivations of the above-referred image and non-image
parameters, after folding in the TACTIC instrumentation details into the
simulated data-bases.  
The results and implications thereof are presented in the next section.

\section{TACTIC Array}  
\begin{figure}[ht]
\begin{center}
\vspace*{0.4cm}
\hspace*{-0.5cm}
\epsfig{file=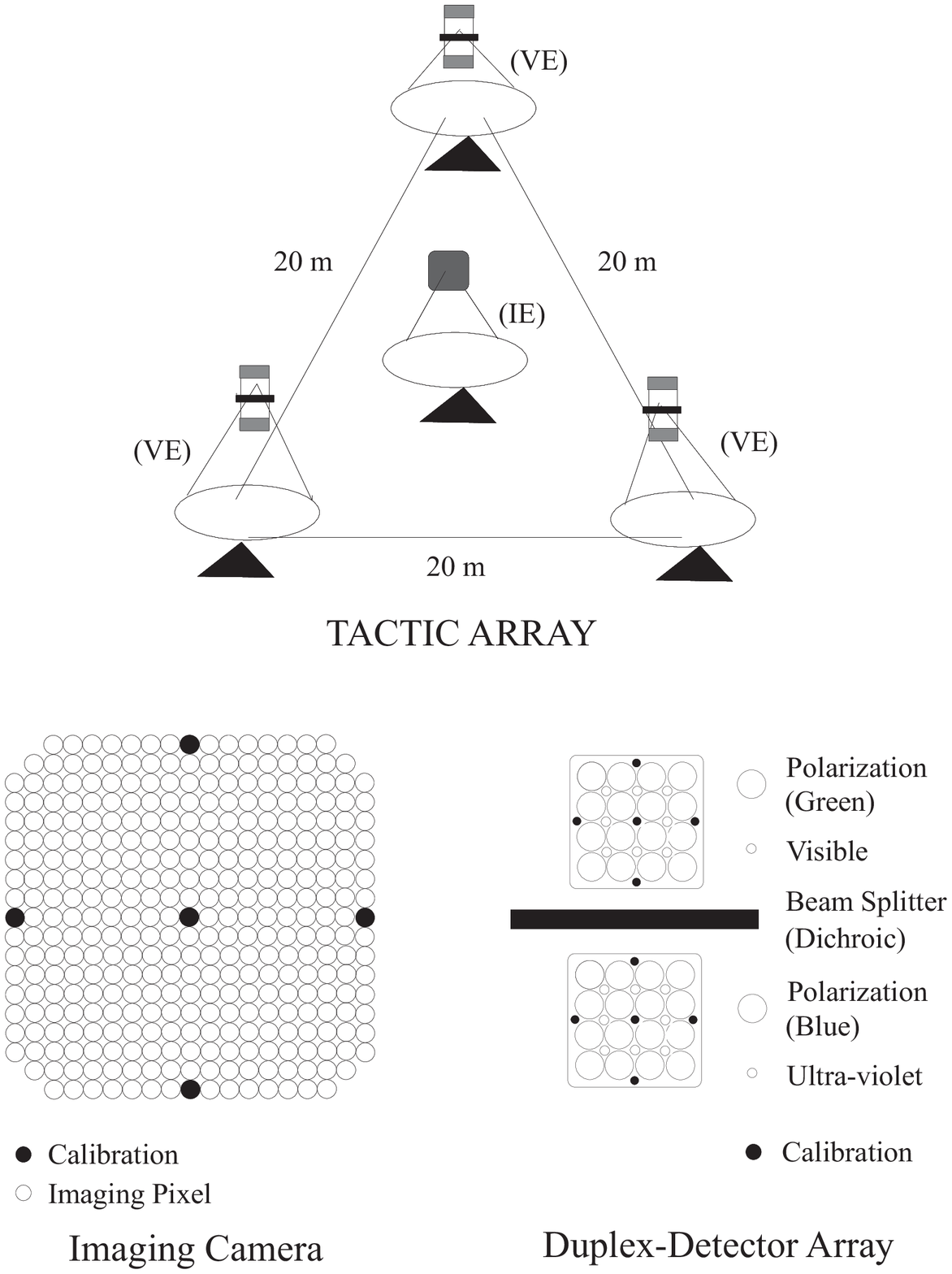,width=12.5cm}
\vspace*{0.4cm}
\caption{\it Sketch of the TACTIC array and the cameras 
for high-sensitivity spectral and temporal investigations on
$\gamma$-ray sources in the photon energy range of $\sim 1-100\,$TeV 
(details see text).}
\label{fig1}
\end{center}
\end{figure}

Experimental details of this instrument, recently  commissioned  at Mt. Abu
($24.62^\circ$ N, $72.75^\circ E$, 1257 m asl)
in the Western Indian state of Rajasthan have been discussed elsewhere
\cite{tactic,Bhat}; we
present here mainly its salient features as relevant to the 
present work: TACTIC
(for TeV Atmospheric Cherenkov Telescope with Imaging Camera) comprises
a compact array of 4 Cherenkov telescope elements, with 1 element (Imaging
Element, IE) disposed at the centroid and 3 elements (Vertex Elements, VE)
placed at the vertices of an equilateral triangle of 20 m 
side (Fig.~\ref{fig1}).
Each telescope element deploys a tessellated light reflector of 
9.5 m$^2$ mirror area
and  composed of 34~$\times$~0.6~m diameter spherical glass mirrors 
(front-aluminized) of a radius of curvature $\sim$ 8 m.
The mirror facets are arranged in a Davis-Cotton geometrical
configuration to yield an on-axis spot-size of 0.2 cm diameter.  The light
reflectors are placed on alt-azimuth mounts and can be synchronously steered
with a PC-based drive system with a source-tracking 
accuracy of $\le$ 2 arcmin. \\
The IE of the TACTIC array yields high-definition images of
atmospheric Cherenkov events generated by $\gamma$-ray and cosmic ray
progenitors with primary energies $\ge$ 1 TeV.  For this purpose, this
element is provided with a focal-plane imaging camera consisting of 349
fast-photomultiplier pixels (Electron Tubes Ltd., 9083UVB) and arranged
in a closely-packed truncated square configuration to yield an FoV
$\sim 6^\circ \times 6^\circ$ with a uniform pixel-resolution 
of $\sim 0.31^\circ$.
The focal-plane instrumentation of each VE is designed primarily to record
various non-image characteristics of the recorded 
Cherenkov event, including its linear polarization state, time-profile and
ultraviolet-to-visible light spectral content (U/V ratio).  For this
purpose, as shown in Figure~\ref{fig1}, this instrumentation consists of
duplex PMT detector (ETL9954 B) arrays which are
placed across a 1:1 optical beam splitter, mounted midway between
the PMT detector arrays at right angles
to the telescope principal axis.
There are $16 \times 0.91^\circ$
diameter PMT detectors in each array which are provided with suitably-oriented
sheet-polarizers (near-UV transmitting, Polaroid-make HNB' type) to measure
the linear polarization state of the recorded Cherenkov events.  In addition,
they are provided with appropriate back-end electronics to record the
time-profile of the detected event with $\sim$ 1 ns resolution.
The response function of this instrumentation, from PMT onwards, can be
approximately represented by a triangular pulse with a rise-time of $\sim$
2 ns and decay time of $\sim$ 4.5 ns.  As is evident from
Figure~\ref{fig1}, a total of 8 PMT pairs are also provided in the VE
focal-plane instrumentation to 'sample' the relative ultraviolet (U-pixels,
$\lambda \sim$ 280-310~nm) to visible (V-pixels, $\lambda \sim$ 310-450 nm)
light content of the recorded events.  Each U-pixel consists of the PMT
type ETL D921 and is placed on the reflection side of the beam-splitter,
exactly opposite to its mating V-pixel (ETL 9097B), placed across
the beam-splitter (transmission side). \\
The TACTIC event trigger is based on a memory-based proximity or
`topological' trigger-generation scheme which stipulates coincident
`firing' of Nearest-Neighbor Non-Collinear Triplet (3 NCT) pixels with
a coincidence resolving gate of $\sim$ 20 ns from within the innermost 240
(16 $\times$ 15) pixels of the 349-pixel IE camera \cite{Bhat}.  
In response to each
such trigger, the TACTIC data acquisition system records the photo-electron
content of each IE and VE pixel (for image, polarization and U/V spectral
information) and the composite time-profile of the Cherenkov event as 'seen'
by all the 32 $\times$ 0.9$^\circ$ dia pixels of the VE camera.  All the PMT
of the IE and VE can be gain-calibrated in an absolute way, based on a scheme
employing the single-photoelectron counting procedure along with radio-active
(Am-241) and fluorescent light pulsers. \\
In the normal (low threshold) mode of operation, deployed
on dark, clear nights, the TACTIC follows a
putative $\gamma$-ray source across
the sky over the zenith-angle range $\theta \le 40^\circ$ and records
background cosmic ray events with rates $\le$ 10 Hz (trigger threshold for
$\gamma$-rays, $\sim$ 0.7 TeV) \cite{bhat}.  On the other hand,
in the supplementary
mode of operation, for UHE $\gamma$-ray astronomy and
cosmic ray composition studies in 10's of TeV energy  region
and telescope array is planned to be employed,
over $\theta$ range $\sim 40- 60^\circ$. This mode of
operation has  the advantage of significantly increasing
the effective detection area and, in turn, leads to an
appreciable enhancement in the relative  number of UHE events coming
from larger core distances and also not suffering from the problem
of image saturation  effects.\\

\section{Simulated data-bases} 
The scheme adopted here for generating TACTIC-compatible
data-bases through the CORSIKA simulation
code (Version 4.5) \cite{capdevielle,heck} using the VENUS code for 
the hadronic high-energy interactions \cite{venus}, 
has been discussed in detail in \cite{haungs1}.  In essence, this
data-base consists of arrival direction and arrival time of Cherenkov photons,
received within a specified wavelength interval by each of the TACTIC mirrors
in response to the incidence on the atmosphere of a $\gamma$-ray photon 
(energy $E_\gamma$ = 50 TeV,
zenith angle $\theta$ = 40$^\circ$) or a cosmic-ray primary (type: proton,
neon or iron nucleus; $E_h$ = 100 TeV, $\theta$ = $40^\circ \pm 2^\circ$).  
Atmospheric
extinction is duly accounted for as a function of 
the Cherenkov photon wavelength.
The received Cherenkov photons are ray-traced into the TACTIC focal-plane
instrumentation and converted into equivalent photoelectrons (pe) after
folding in the mirror optical characteristics \cite{rannot}, 
the photocathode spectral
response and the beam-splitter wavelength-dependent reflection and transmission
coefficients (last one in case of VE only).  The 2D-distribution of the number
of pe, thus registered by various pixels of the IE camera, constitutes the
high-resolution Cherenkov image.  Similarly, the pe contents, separately
registered by all the U- and V- pixels of a VE camera, yield the U/V 
spectral ratio of the Cherenkov event, while the 'arrival-time' distribution of
the pe's, as noted in the image-plane of a VE camera 
$(32 \times 0.9^\circ$ dia
PMT only), gives the time-profile of this event.  The sheet polarizers, 
normally used with these 32 PMT of the VE camera for 
polarization measurements, 
are assumed to be absent for the present simulation exercise.  100 showers 
each of $\gamma$-rays (50~TeV) and protons, neon and iron nuclei (100~TeV) have
been simulated in this manner. \\
This scheme for data-base generation allows to simultaneously
obtain the desired data for the TACTIC array, assumedly placed 
at various distances
(R $\sim$ 10~m - 300~m) from the shower core.  We use here the 
data corresponding
to the representative  R $\sim$ 195~m, consisting of only 100 Cherenkov images
and associated time-profile and U/V data and belonging to each of the 4 
primary types considered here. Similar data has been obtained for  two other core distances R $\sim$ 245~m and R $\sim$ 295~m.

\section{Hillas Parameters} 
Cherenkov images of $\gamma$-ray
showers are mainly elliptical in shape, hence compact.  The Cherenkov images
of hadronic showers are mostly irregular in shape, implying that
gamma rays can be distinguished from hadron events on the basis of shape
and compactness of Cherenkov images.  The process of $\gamma$/hadron
separation needs parameterization and this parameterization was first
introduced by Hillas \cite{hillas}. The resulting Hillas parameters 
can be generally classified into either `shape' parameters such as Length (L)
and Width (W) which characterize the size of the image, or into orientation
parameters such as alpha $\alpha$, which is the angle of the
image length with the direction of the source location within the
field of view of the camera. The azwidth parameter combines both
the image shape and orientation features.\\
We have calculated the following second moments
for the simulated Cherenkov images: shape parameters like
Length (L), Width (W) and Distance (D) and  orientation parameters like
alpha ($\alpha$), Azwidth (A), and Miss (M). 
In so far as the image orientation parameters are concerned, 
they are found to be small for $\gamma$-rays, assumed here to be coming
from a point-source placed along the telescope-axis. For the 
nuclear progenitors,
$\alpha$ (or other orientation parameters) has no significant classification
potential, since all these particles are assumed to be randomly oriented around
the telescope axis (the same conclusion would hold for $\gamma$-rays of a truly
diffuse origin). The average values of some representative  Hillas
parameters are listed in Table~\ref{tab1} for comparison,
where the different hadronic primaries are combined. 
\vspace*{0.5cm}
\begin{table}[ht]
\caption{\it Average values of Hillas parameters for 
primary gammas and hadrons. The values of the different
hadron species are combined. All values are in angular degrees.}
\label{tab1}
\vspace*{0.2cm}
{\small
\begin{tabular}{|l|c|c|c|c|c|c|}
\hline
Parameter&Length&Width&Distance&Azwidth&Miss&Alpha\\ \hline
Gamma rays&0.62&0.42&1.0&1.4&0.18&6.4 \\
\hline
hadrons&   0.71&0.43&1.2&0.56&0.53&32.8 \\
\hline
\end{tabular}
}
\vspace*{0.4cm}
\end{table}

\section{Multifractal moments}
Employing the prescription given in \cite{haungs1},
we have calculated the multifractal moments \cite{mandelbrot,aharony} 
of the simulated TACTIC images
recorded within the innermost 256 pixels of its imaging camera.
For this purpose, the
image has been divided into $M = 2^{\nu} $ equal parts,
where $\nu$ =2,4,6 and 8 is the chosen
scale-length.   
The multifractal moments are given as:
\begin{equation}
G_q (M) = \Sigma^M_{j=1} (\frac {k_j} {N})^q
\end{equation}
\noindent where N is the total number of pe in the image,
$k_j$ is the number of pe in the k$^{th}$ cell and q is the order of the
fractal moment.  In case of a fractal, $G_q$ shows a power-law behavior
with M, i.e.,
\begin{equation}
G_q \sim M^{\tau_q}
\end{equation}
where
\begin{equation}
\tau_q = \frac {1} {ln 2} \frac {d ~ ln ~ G_q} {d \nu} \,\, .
\end{equation}
For a fractal structure, there exists a linear
relationship between the natural logarithm of $G_q$ and $\nu$  and the slope
of this line, $\tau_q$, is related to the generalized multifractal
dimensions, $D_q$, by:
\begin{equation}
D_q = \frac {\tau_q} {q-1}, q \neq 1 
\end{equation}
where  -6 $<$ q $<$ 6 is the order the multifractal moment.
In \cite{haungs1}, we have used 2 multifractal dimensions, 
$D_2$ and $D_6$, as classifiers
and have achieved a fairly high discrimination power through them for a large
composite database of 2,4000 images. 
On the contrary, since the database
size here is significantly smaller (only 100 images for each primary
type), it is 
difficult in this case to clearly identify one (or few) fractal dimensions
by the method of overlap of distributions. \\
By using the correlation ratio we were 
able to identify $D_4$  and $D_6$
as having a minimum correlation. 
A correlation ratio between two variables,
say X and Y, is defined as
\begin{equation}
r=\frac{\sum_{i=1}^{n}(X_i - \overline{X})(Y_i -\overline{Y})}
    {\sqrt{(\sum_{i=1}^{n}(X_i -\overline{X})^2)
           (\sum_{i=1}^{n}(Y_i -\overline{Y})^2)}}
\end{equation}
The correlation ratio r is a measure of linear association between
two variables. A positive coefficient indicates that, as one value increases
the other tends to increase whereas a negative coefficient indicates
as one variable increases the other tends to decrease.
$D_4$  and $D_6$ are thus relatively independent
and should as such provide the best possible segregation of the primary
masses.

\section{Wavelet Moments}  
Wavelets can detect both the location and
the scale of a structure in an image.  These are parameterised by a scale
(dilation parameter) 'a' $>$ 0 and a translation parameter 'b' (-$\infty <$
b $< \infty$) \cite{debauchies,greiner}, such that
\begin{equation}
\phi (x) = \frac {\psi(x -b)}{a}\,\,.
\end{equation}
\noindent Since we are analyzing Cherenkov images which are fractal in nature,
it is the dilation parameter 'a' which is of interest to us here rather
than the translation parameter 'b'.  The wavelet moment \cite{debauchies} 
$W_q$ is given as :
\begin{equation}
W_q (M) = \Sigma^M_{j-1} (\frac {| k_{j-1} - k_j |} {N})^q
\end{equation}
\noindent where $k_j$ is the number of pe in the j$^{\rm th}$ 
cell in a particular scale, and
$k_{j+1}$, in the j$^{\rm th}$ cell in the consecutive scale.  
The wavelet moment
has been obtained by dividing the Cherenkov image into M = 4, 16, 64 and 256
equally-sized parts with 64, 16, 4 and 1 PMT pixels respectively and
counting the number of pe in each part.  The difference of probability
in each scale gives the wavelet moment.  It turns out that for a Cherenkov
image
\begin{equation}
W_q \sim M^{\beta_q}
\end{equation}
\noindent implying that the wavelet moment $W_q$ bears a power-law 
relationship with M.
As was shown in \cite{haungs1}, the exponent $\beta_q$ (q = 1-6) is found to be
sensitive to the structure of the Cherenkov image and has the lowest value
for $\gamma$-rays, followed by protons, neon and iron nuclei, in increasing
order of the nuclear charge. In \cite{haungs1} only the wavelet parameters
$\beta_2$ and $\beta_6$ were used for image characterization on account
of the fairly large database deployed there. In the present work, we 
prefer to use the wavelet parameters $\beta_1$ and $\beta_5$ as
classifiers. They have been chosen by the method of minimum correlation.

\section{Time parameters}  
\begin{figure}[ht]
\begin{center}
\vspace*{.3cm}
\hspace*{-.2cm}
\begin{tabular}{cc}
\epsfig{figure=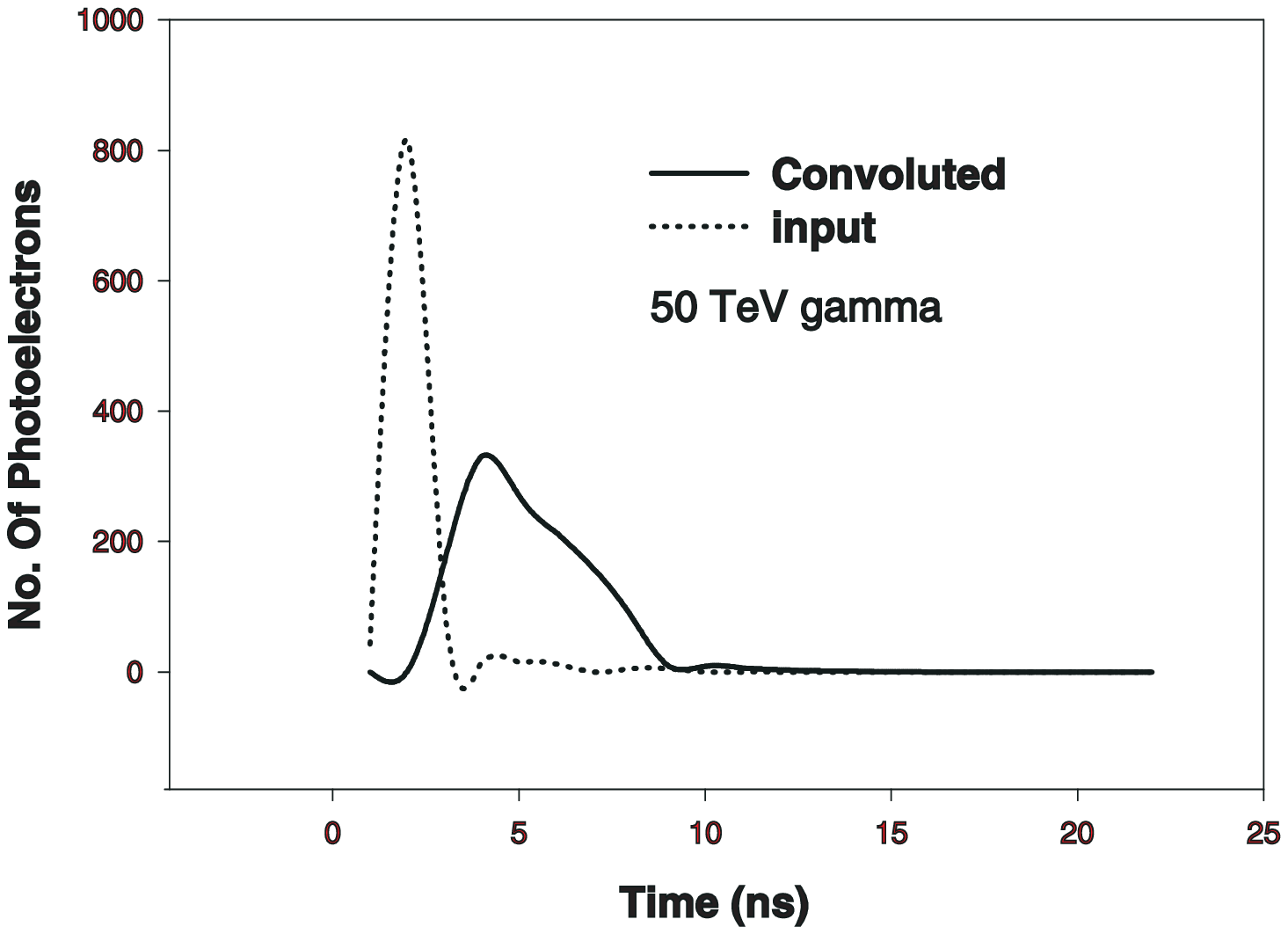,clip=,width=6.6cm} & 
\epsfig{figure=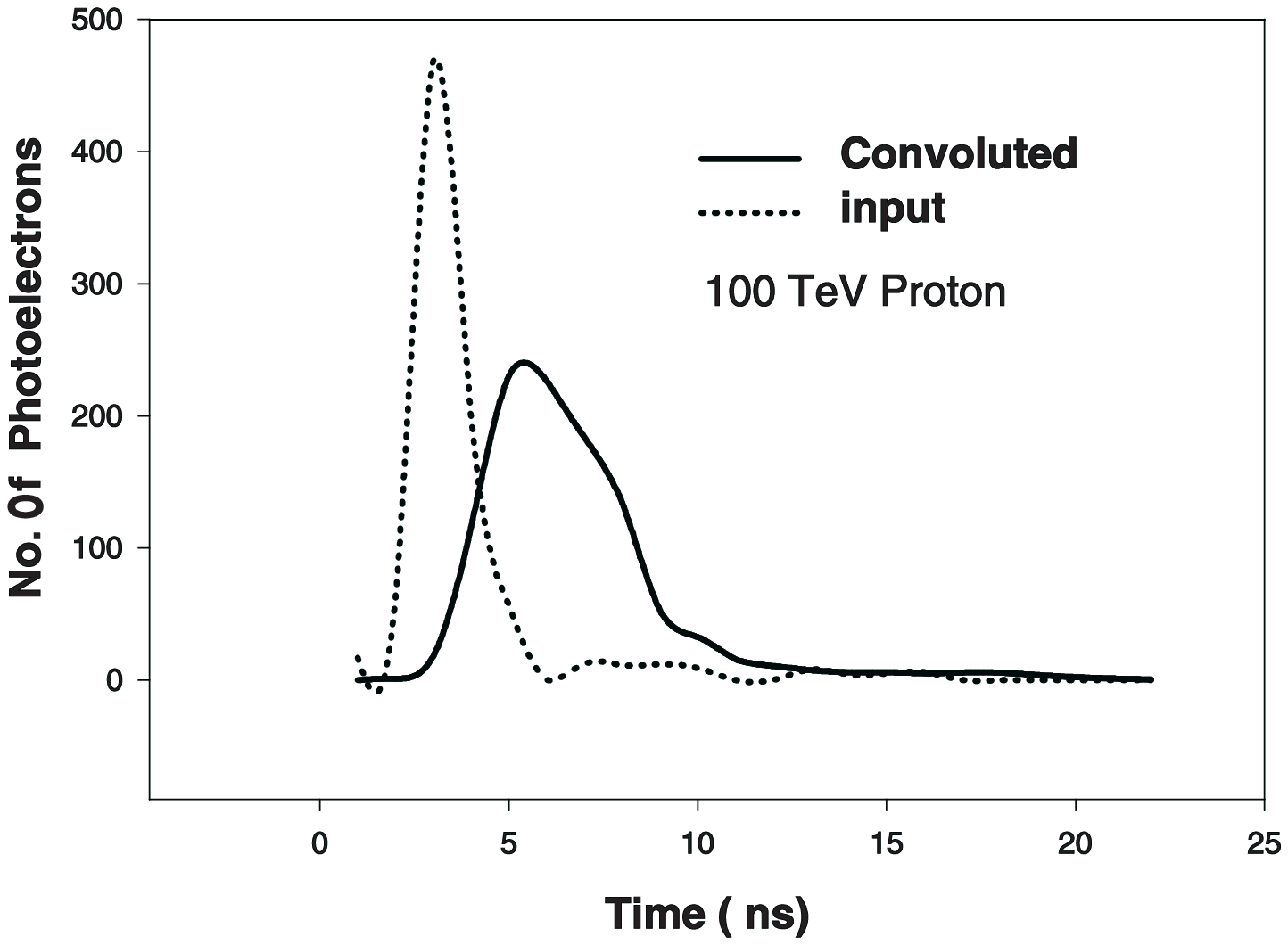,clip=,width=6.6cm} \\ 
\epsfig{figure=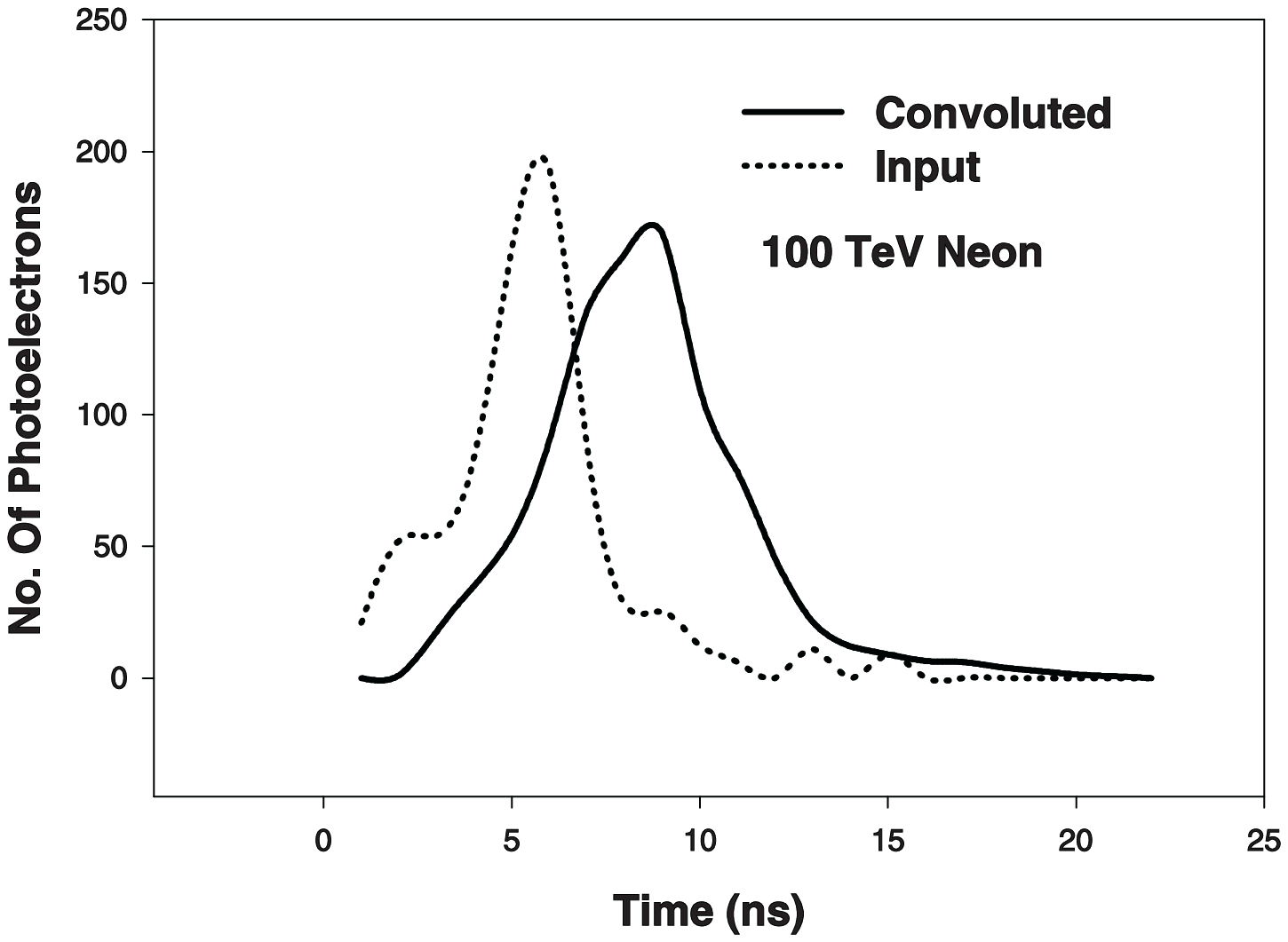,clip=,width=6.6cm} & 
\epsfig{figure=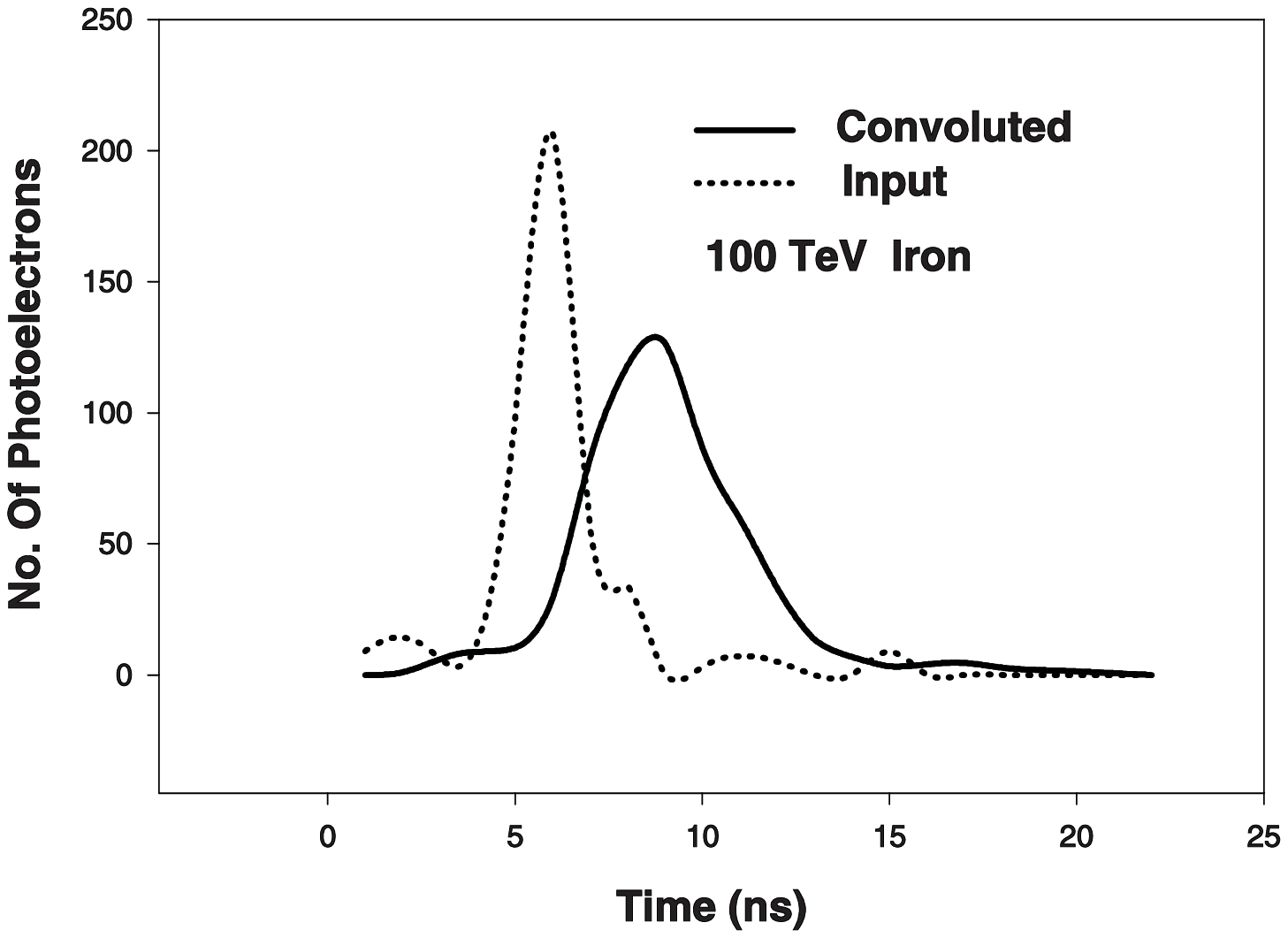,clip=,width=6.6cm} \\ 
\end{tabular}
\caption{\it Pre-detector (input) and post-detector (convoluted) Cherenkov
arrival time profiles for $\gamma$-ray, proton, neon and iron initiated
showers.}
\label{fig2} 
\vspace*{.5cm}
\end{center}
\end{figure}

There have been several reports in literature
tentatively suggesting a dependence of various time-parameters of
Cherenkov pulse-profiles, including their rise-times and base-width on the
progenitor type.  While most of these suggestions are based on
simulation studies \cite{rodriguez}, there is one piece of 
experimental work wherein
temporal profiles of Cherenkov pulses, recorded by an atmospheric Cherenkov
telescope system, have been utilized to preferentially select $\gamma$-ray
events from the cosmic-ray background events.  Using a largely ad-hoc
approach for this selection, this group claimed the detection of a
TeV $\gamma$-ray signal from the Crab Nebula in 1.5 hours of
observation at 4.35 significance level. The observations were
carried out with a 11 m-diameter solar collector-based
Cherenkov telescope \cite{tumer}. More recent simulation
investigations \cite{aharonian} have indicated that the 
differences in the Cherenkov pulse
profiles of different primary types are related to the various details of
development of EAS initiated by $\gamma$-ray and nuclear-primaries in the
atmosphere, including the important role of muon secondaries in the 
latter case.
Thus, the pulse from a $\gamma$-ray primary is expected to have a relatively
smooth profile with a typical rise-time of $\sim$ 1 ns, and a decay-time
of $\sim$ 2 ns, while the temporal profile of a hadron-origin has
relatively longer rise and decay times and, in addition, a superimposed
microstructure, possibly due to Cherenkov light produced by single muons,
moving close to the detector system. \\
One of the output parameters in the CORSIKA is the Cherenkov photon
arrival time. The  measurement of time begins with the
first interaction and time taken by
each photon as it traverses the atmosphere and reaches
observation level to the focal plane of TACTIC is measured.
Figure~\ref{fig2} represents CORSIKA-generated typical waveforms 
expected for the TACTIC for the 4 progenitor
types considered here: $\gamma$-ray (50 TeV), proton (100 TeV), neon (100
TeV) and iron (100 TeV).  Both the cases have been displayed: time-profiles
expected at the input of the TACTIC focal-plane instrumentation and also
at the back-end of this instrumentation (amplifier output), after
convoluting the input time-profiles with the expected response function of
the TACTIC instrumentation -- reasonably approximated by
a triangular pulse with a rise-time
of 2 ns and fall time of 4.5 ns.  With other authors confining themselves
largely to analyzing pulse-profiles at the detector input stage only, the
results from the exercise, carried out  here, should
be of more practical importance, for they include effects that need be
considered, like the loss of the time profile fine-structure, due to the
relatively slow response-time of typical Cherenkov telescope systems, like
the TACTIC. \\
It is important to study the shape of the Cherenkov photon  arrival time
distribution as represented by some pulse profile. So by fitting a suitable
probability function, it is possible to parameterise the arrival time
distribution. In most of the previous studies, generally,
fittings have been done  only at
pre-detector state  by using distributions like $\Gamma$-function and
Lognormal functions. In the present study  arrival-time
distribution of the pe, obtained at the post-detector stage for
each simulated event, has been fitted with exponentially modified Gaussian
(emg) \cite{2d manual} and half-Gaussian modified Gaussian (gmg) 
\cite{2d manual} functions of the following
forms, respectively:
\vskip 0.2cm
emg functions:
\begin{equation}
f(t) = {\frac{ac \sqrt{2\pi}}{2d}}\exp[\frac{b-t}{d}+
\frac{c^2}{2d^2}][\frac{d}{|d|}-{\rm erf}(z1)] 
\end{equation}
\vskip 0.2cm
where
\begin{equation}
z1=\frac{b-t}{\sqrt{2}c}+\frac{c}{\sqrt{2}d}
\end{equation}
\vskip 0.2cm
gmg function :  
\begin{equation}
f(t) = {\frac{ac}{\sqrt{d^2+c^2}}}\exp[-\frac{1}{2}\frac{(t-b)^2}
{d^2+c^2}][1 +{\rm erf}(z)]
\end{equation}
where
\begin{equation}
z =\frac{1}{2}\frac{\sqrt{2}d(t-b)}{c\sqrt{(d^2 + c^2)}}
\end{equation}
\noindent The various properties of  these two functions are
Amplitude = $a$,
Centre    = $b$,
area  = $\sqrt{2 \pi}ac$,
FWHM = $2\sqrt{2ln2}c$,
and time constant = $d$. \\
Both, emg and gmg functions have
two small constraints, viz., c $>$ 0, and d $\ne$ 0.
Rise time and base width ($t_{90}$ -$t_{10}$) which are obtained for each
convoluted pulse profile has been determined. 
\vspace*{0.5cm}
\begin{table}[hb]
\caption{\it Average values of rise time
and base width for gamma rays and hadrons in nanoseconds. The event 
distributions are shown in Fig.~\ref{fig3}.}
\label{tab2}
\vspace*{0.2cm}
{\small
\begin{tabular}{|l|cccc|}
\hline
&gamma rays&protons&neon Nuclei&iron Nuclei\\
\hline
Rise Time&1.82&2.71&2.94&3.36\\
\hline
Base Width&7.29&8.50&9.37&10.06\\
\hline
\end{tabular}
}
\vspace*{0.4cm}
\end{table}
While rise time as defined here is the time between
10$\%$ and 90$\%$ of the peak value, the base width is the time elapsed
between 10$\%$ of peak value on both sides of the peak.
We have used Table curve 2D (Jandel Scientific Software) \cite{2d manual} for
fitting the above-referred two functions.   
Table~\ref{tab2} lists the mean values and the overall ranges of these 
\begin{figure}[ht]
\begin{center}
\vspace*{.2cm}
\hspace*{-.8cm}
\begin{tabular}{cc}
\epsfig{figure=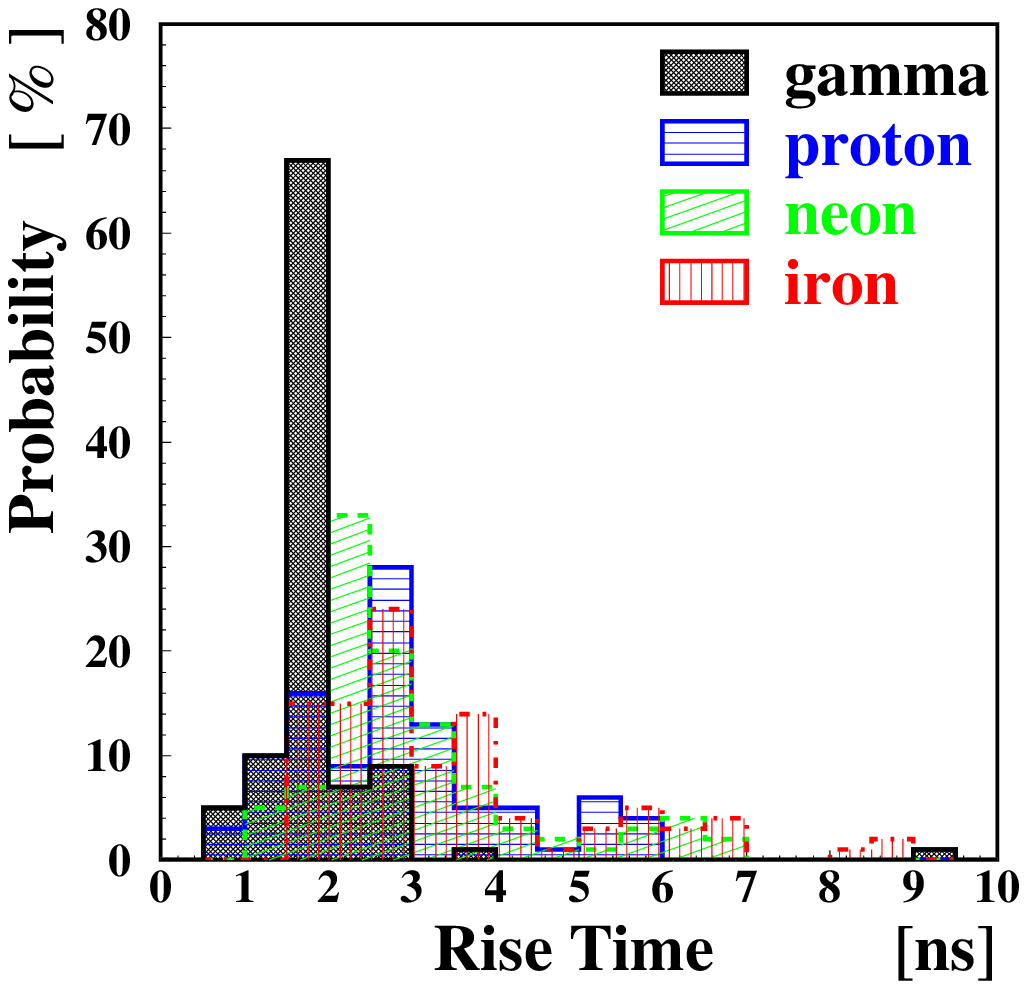,width=7.2cm} &
\epsfig{figure=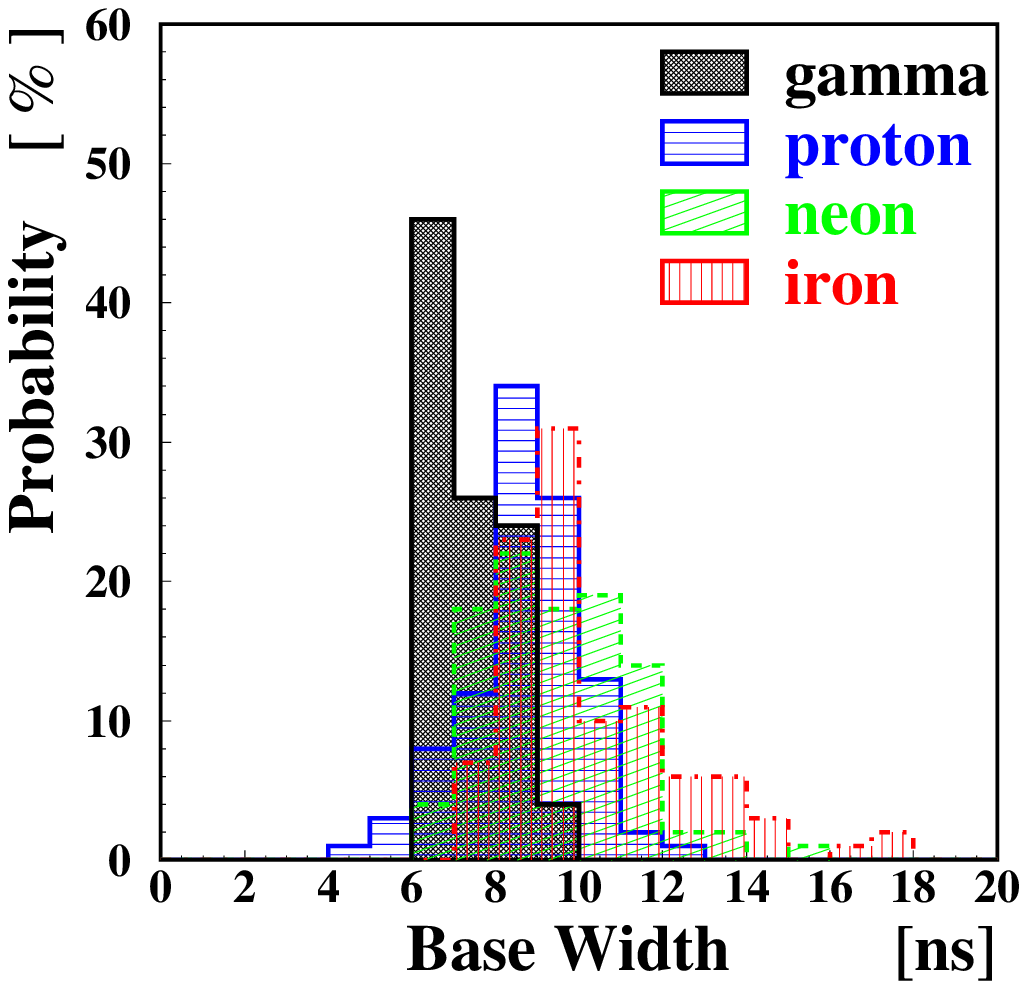,width=7.2cm} \\
\end{tabular}
\caption{\it Rise time and base width parameters of Cherenkov arrival
time pulse profiles (post-detector) for $\gamma$-ray, proton, neon and
iron initiated showers.}
\label{fig3} 
\vspace*{.5cm}
\end{center}
\end{figure}
parameters as derived for
the simulated data. It is evident from both 
Figure~\ref{fig2} and Table~\ref{tab2}
that $\gamma$-ray images have relatively smaller mean values, 
while iron nuclei have the largest values for these parameters,
followed by neon nuclei and protons. Figure~\ref{fig3} depicts the
distributions of rise time and
base width for all species considered here.

\section{U/V Spectral Ratio} 
Attention was first drawn by the Crimean Astrophysical
Observatory group \cite{stepanian} towards the possible diagnostic role of
the Ultraviolet (U; $\lambda \sim$ 250-310 nm) content of a Cherenkov event 
compared with the corresponding
photon yield in the Visible (V; $\lambda \sim$ 310-500 nm) region -- U/V 
spectral ratio -- for
differentiating between $\gamma$-rays and hadrons.  The underlying rationale
for this expectation is that, in case of $\gamma$-rays showers, 
the bulk of Cherenkov
light is produced at relatively high altitudes ($\ge$ 8 km) by 
electron secondaries
and the resulting light will be relatively deficient in the U-component because
of rather strong atmospheric extinction effects for ultraviolet 
radiation coming
from these altitudes.  On the contrary, hadron showers are accompanied by
relativistic muons which penetrate down to lower altitudes and generate
relatively U-richer Cherenkov light closer to the observation plane.  
Different
groups have expressed different and sometimes contradictory views about
the efficacy of this parameter for event characterization purpose.  
The exact
value for the U/V parameter for a given progenitor species is evidently a
function of the 
\begin{wrapfigure}[20]{l}{7.4cm}
\vspace*{0.cm}
%\begin{center}
\epsfig{figure=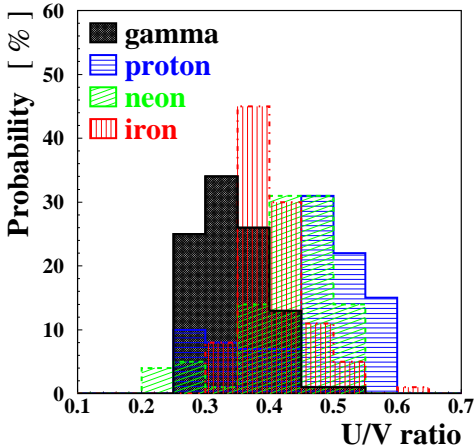,width=7.2cm} 
\vspace*{-0.4cm}
\caption{\it U/V ratio of $\gamma$-ray, proton, neon, and iron 
initiated showers
at a typical distance of 195 m from the core.}
\label{fig4} 
\vspace*{0.5cm}
%\end{center}
\end{wrapfigure}
actual detector configuration and spectral response as
also the exact atmospheric conditions, apart from the effects of 
shower-to-shower
fluctuations, etc. For practical reasons, the geometrical disposition of
the U- and V- channel PMT and the detector size in the TACTIC instrument
are not particularly favorable for measurements of the
U/V parameters. Nevertheless,
it is evident from Figure~\ref{fig4} that the expected distributions of
this parameter for the TACTIC detector configuration  are not completely
overlapping and may as such help along with other parameters in separating
$\gamma$-ray from hadrons.

\section{ANN Studies}
We have examined the implied classification potential of the above referred
parameters more quantitatively by appealing to the
pattern-recognition capabilities of an artificial neural network (ANN). \\
We have performed ANN 
studies by using the Jetnet 3.0 ANN package~\cite{lblad} with two 
Hillas parameters (W and D),
two fractal dimensions ($D_4$, $D_6$), two wavelet moments
($\beta_1$, $\beta_5$), two time parameters (rise time, base width), 
and the U/V parameter as  input observables.
%We choose two Hillas parameters Width (W) and Distance (D) by the
%method of minimum correlation.
All the three hadron species have been taken together
as a single cosmic-ray family. Thus the overall data-base used here 
consists of two types of events:
100 $\gamma$-rays and 300 hadrons, the
latter consisting of proton, neon and iron nuclei taken in equal proportions.
One half of the events have been used for training the ANN and the other 
half, for independently testing the trained net.  
For testing the stability of the net the training-testing procedures are
repeated several times by varying the events belonging to the two samples.
The output values demanded are
0.0 for $\gamma$-rays and 1.0 for hadrons. 
In all the ANN results discussed below, we have chosen %
\begin{wrapfigure}[20]{l}{7.4cm}
\vspace*{-0.4cm}
\epsfig{figure=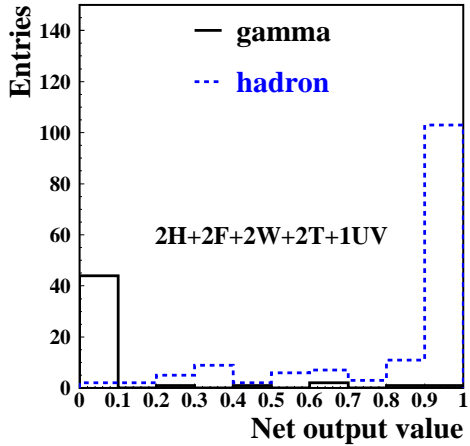,width=7.2cm} 
\caption{\it Event distribution of the neural net output of $\gamma$-rays 
and hadrons using all parameters corresponding to impact points of
195~m.}
\label{fig5} 
\vspace*{0.5cm}
\end{wrapfigure}
the net sigmoid
function as 
$g(x) = 1/ (\exp{-2x} + 1)$, 
with two hidden layers, one with 25 nodes and the other with 15
nodes. We have used the ANN in the back-propagation mode  with
a learning rate $\eta$=0.05 and
the momentum parameter $\alpha$ = 0.5.
For optimizing the cut value and checking the net quality 
there is a parameter called
separability-index G \cite{roth}, defined as \\
\hspace*{2cm} $G=\sqrt{\prod_{i=1}^{n} P_i}$ \\
where n is the number of output classes (in our case, n=2) and $P_i$
denotes the probability of an event of class i to be classified
in the right class.
The maximum value of the G-index is 1.0 for an ideal
segregation of the two classes.
We calculate $G_1$ and $G_2$ which are the G-indices 
corresponding to the data of the training and testing samples, respectively.\\ 
For the examination of the separation quality of the net,
we have trained the net with different parameters and their
combinations of the mass
sensitive observables (see Table~\ref{tab-g}).
A multiparameter analysis method like neural net 
is able to find the best combination of different parameters.
But if more parameters are used in the analysis the statistics
of the Monte Carlo simulated events, necessary for  
training of the net, have to be increased
in proportion (curse of dimensionality~\cite{bishop}).
Hence the parameters with the least
correlation have to be considered. The least correlated parameters
from each set, i.e fractals, wavelets, Hillas-parameters etc., 
have been identified
and results are given in Table~~\ref{tab-g}. The similarity of the
values of $G_1$ and $G_2$ indices indicates proper functioning of the net.
The quality factor Q (ratio of $\gamma$-selection efficiency
and the square root of the background selection efficiency)
has been calculated for each case and corresponding
values are given in the table which indicates that improved results
are obtained for the case when imaging and non-imaging 
parameters are used together.
The result of this last example is shown in Fig.~\ref{fig5}. 
Similar ANN exercise has been repeated with all
above mentioned parameters (except time due to non-availability of data)
for data sets corresponding to core-distances of 245~m and 295~m.
The results obtained are consistent 
with the corresponding results obtained at 195~m.
From Table~\ref{tab-g} it is clear that fractal parameters are better
for $\gamma$/hadron separation than wavelet parameters
for this small database. It is also evident that two time parameters
(rise time and base width) show good segregation between $\gamma$-rays and
hadrons and become more significant when combined with Hillas- , fractal- 
and wavelet parameters. 
On the other hand the U/V-ratio is a poor segregator 
but when used in combination with parameters of the other families
it helps to improve the segregation quality. This
behaviour is seen for all the three core distances.
\vspace*{0.01cm}
\begin{table}[h]
\caption{\it Details of the ANN studies. (--) means indeterminate values.
The statistical uncertainties are in the order of $20\%$.}
\label{tab-g} 
\vspace*{0.2cm}
\begin{small}
\begin{tabular}{|cccc|}
\hline
Set of parameters&\multicolumn{2}{c}{Separability-index}&Quality factor\\
 & $G_2$ & $G_1$  &        Q \\
\hline
2F&              0.72&0.77&1.6\\
2W&              0.66&0.66&--\\
2T&              0.85&0.87&1.1\\
1UV&             0.53&0.60&--\\
2H&              0.65&0.70&--\\
2H+1UV&          0.83&0.65&0.7\\
2H+2T&           0.84&0.87&1.5\\
2H+2F&           0.74&0.77&1.9\\
2H+2W&           0.74&0.69&0.4\\
2H+2T+1UV&       0.88&0.74&2.4\\
2H+2F+2T+2W&     0.89&0.89&3.0\\
2H+2F+2T+2W+1UV& 0.84&0.85&3.4\\
\hline
\end{tabular}
\end{small}
\vspace*{-0.2cm}
\end{table}

\section{Discussion and Conclusions}
The present exploratory study is remarkable for its
deliberate choice of extremely small data-bases of $\gamma$-ray and
hadron events and for seeking their efficient segregation through deployment
of an unusually large number of parameters characterizing these events.  An
artificial neural network, having intrinsically superior and fault-tolerant
cognitive capabilities and utilizing correlative features underlying 
these parameters for the net training and testing, is employed for achieving
the desired classification.  While as a sufficiently large data-base may
help to properly train a neural net for $\gamma$-ray/hadron classification,
based on only a limited number of event parameters, as demonstrated in
the case of Hillas parameters by some previous authors, it is not expected
to do so efficiently when the data-base size is restricted, as 
can happen in various
practical situations.  The latter view-point is clearly endorsed by the results
presented in the first part of the previous section for various individual
parameter families separately.
The situation is found to improve favorably when several
parameter families are used together and their underlying complementary
properties exploited.  The results presented here are remarkable in one
respect, viz, $\gamma$-ray events are accepted at $>$ 90 $\%$ level,
as compared to ($30 - 50\%$) $\gamma$-ray
acceptance levels presently available through Hillas and supercuts image
filter strategies.  
This significantly higher results of 
acceptance levels for $\gamma$-rays provided by the multiparameter approach,
despite the extremely small data-base size, is of practical importance
in ground-based $\gamma$-ray astronomy for the following two important
reasons: The $\gamma$-ray source spectrum can be inferred with less
ambiguity and/or lower-flux signals can be retrieved more efficiently.
Another interesting observation made from the present study is
that, since the ANN does not need large volumes of data ($\gamma$-rays or
hadrons) for training when a sufficiently multi-dimensional parameter
space is available, the necessary training can in principle, be imparted
with actual (rather than simulated) $\gamma$-ray data, which any moderately
sensitive experiment can record over a reasonable length of time from
a known cosmic source, e.g., Crab Nebula, Mkn 501 or Mkn 421. \\
The following 'caveats' need to be kept in mind in the context
of the present exploratory work: 
Same primary energy (50 TeV for $\gamma$-rays and 100
TeV for the hadrons) has been considered. 
It would  be in order to make a more realistic investigation involving
ultra-high energy $\gamma$-rays and hadrons drawn from a typical spectrum
and range of core distances going all the way to 500 m.
Another related  activity would be  to seek the use of this 
enlarged parameter base for cosmic-ray
mass composition studies as an extension of \cite{haungs1} 
and the present work
and make it more definitive for handling real data going to be generated by
the TACTIC array of Cherenkov telescopes.  These studies are presently in
progress and the results would be presented elsewhere. 

{\ack This work has been supported by the Indian-German 
bilateral agreement of scientific-technical cooperation 
(WTZ INI-205). The authors gratefully acknowledge the help
received from their colleagues, Mr. M.L. Sapru
and Dr. D. Heck in preparing some of the used simulations, 
and in particular the valuable discussions with Dr. M. Roth
about neural nets.}

\end{document}